\documentclass[10pt]{article}
\usepackage[left=1in,right=1in,top=1in,bottom=1in]{geometry}
\usepackage{url,lineno,microtype,subcaption,color}
\definecolor{niceblue}{RGB}{0,0,100}
\usepackage[colorlinks=true,linkcolor=niceblue,citecolor=niceblue,urlcolor=blue]{hyperref}
\usepackage[onehalfspacing]{setspace}
\usepackage{arydshln}
\usepackage{upgreek}
\usepackage{graphicx}
\usepackage{amsmath}

\usepackage[maxbibnames=99,citestyle=authoryear,style=authoryear,backend=bibtex,hyperref]{biblatex}
\renewbibmacro*{pageref}{%
 \printtext[parens]{%
  \Acrobatmenu{GoBack}{Go back}%
 }%
}

\begin{document}

	\begin{titlepage}
	\title{\LARGE \bf Network Growth Under Opportunistic Attachment}
	\author{Carolina Mattsson\footnote{CENTAI, Corso Inghilterra 3, 10138, Turin, Italy. mattsson.c@northeastern.edu.} \vspace{5mm}}
	\date{\today}
	\maketitle
	
	\begin{abstract}
    \normalsize Growing network models can potentially be a useful tool in the development of economic theory. This work introduces an ``opportunistic attachment'' mechanism where incoming nodes, in deciding where to join a network, consider features of the entry points available to them. For example, an entrepreneur looking to start a thriving business might consider the expected revenue of many hypothetical businesses. This mechanism is explored, in isolation, via a minimal model where PageRank serves to score the available opportunities. Despite its simplicity, this model gives rise to rich node dynamics, path-dependence, and an unexpected degenerate structure. We go on to argue that this model might be useful to theoretical development as a maximally stylised model of entrepreneurial growth. Central to the argument is an alternative set of microfoundations introduced in~\cite{leontief_money-flow_1993} whereby the steady state of a random walk is a notion of economic equilibrium. To the extent this argument holds, our findings suggest that entrepreneurs face a shifting ``opportunity space'' where the number of potential business opportunities is effectively unbounded. Opportunistic attachment is thus a candidate mechanism for relating the structure of an economic system to its future growth.
	\end{abstract}
	\vspace{3cm}
	\setcounter{page}{0}
	\thispagestyle{empty}
	
\end{titlepage}
\pagebreak \newpage

\section{Introduction}

Models of growing networks have proven to be useful in studying social and biological systems, where it is now understood that relatively simple mechanisms can lead to the emergence of complex phenomena. For instance, linear preferential attachment---where incoming nodes are more likely to establish ties to existing nodes with more existing ties---is enough to produce highly non-linear ``rich get richer'' dynamics~\parencite{barabasi_emergence_1999,krapivsky_connectivity_2000,barabasi_network_2016}. This has well-studied implications for the resilience, dynamics, and predictability of complex systems~\parencite{albert_error_2000,pastor-satorras_epidemic_2001,salganik_experimental_2006}. There is also substantial evidence of preferential attachment in economic systems~\parencite{atalay_network_2011,ozaki_integration_2024}. Indeed, economic systems can be seen as growing networks in the sense that new businesses (or, any economic entities) establish ties to existing ones as they begin economic activities. However, it is not clear that preferential attachment would be the only mechanism relevant for growing economic systems.

This work introduces ``opportunistic attachment'' to reflect stylized scenarios where incoming nodes, in deciding where to join the network, consider features of the entry points available to them. As an example, entrepreneurs generally consider the expected revenue of a hypothetical business in deciding whether to start that business~\parencite{cassar_entrepreneur_2006}. The expected revenue depends on both which suppliers are selected and which customers are targeted. The corresponding family of growing network models is one where ``opportunity'' is operationalized as a metric computed for each entry point that a new node can choose to step into. This is substantially different from the family of growing network models where ``preference'' is operationalized as a metric computed for the existing nodes to which a new node can choose to connect~\parencite{overgoor_choosing_2019,overgoor_scaling_2020}.

We study growing networks under opportunistic attachment where PageRank~\parencite{page_pagerank_1999} is used to quantify opportunity. In the minimal version, incoming nodes place one incoming link and one outgoing link. The projected PageRank is computed for each entry point available to the incoming node. In close analogy to generalized preferential attachment~\parencite{krapivsky_connectivity_2000}, incoming nodes select a network position in proportion to this value to a power of $\gamma$. Selection is random when $\gamma = 0$, and, when $\gamma$ is high enough, incoming nodes consistently choose optimally with respect to the opportunity score of the available entry points. This paper focuses on the early behavior of this minimal model, where exhaustive search remains computationally feasible. More generally, incoming nodes encounter a rapidly expanding opportunity space in that the number of possible entry points grows combinatorially with the number of existing nodes. 

Computational simulation reveals that highly endogenous, path-dependent system dynamics emerge already with the minimal model of network growth under opportunistic attachment. Initial node rank is highly stochastic, but the distribution of opportunity scores encountered by later nodes depends on the intensity of selection by earlier nodes. When the selection of entry point is near-optimal, the stochastic entry of a single node can appreciably shift the opportunity space for subsequent nodes. We see familiar patterns of advantage over time for early nodes at lower selection intensities. However, when choices are sufficiently opportunistic we begin to see a second dynamic whereby new nodes have a chance to take advantage of new opportunities and some are lucky enough to become larger than the old. Finally, when taken to the extreme, these dynamics produce network growth in one principal direction where early nodes end up ``left behind''.

It is then argued that this growing network model could be useful as a tool for the development of theory on entrepreneurial growth in economic systems. In support of this argument, we outline a theoretical basis for the main assumptions relating our growing network model to entrepreneurial economies. Central to the argument is an alternative set of microfoundations introduced in~\cite{leontief_money-flow_1993} whereby ``the flow of money [can be] described as a Markov chain. Its ergodic state is equivalent to the economic equilibrium'' (\emph{Theorem 1}, p. 228). This deliberate theoretical grounding makes our growing network model a highly stylized representation of economic systems where ``entrepreneurs'' seeking out profitable business opportunities is the dominant mechanism of growth.

The contribution of this paper is to suggest that several complex phenomena found in real economic systems might be usefully related via an opportunistic attachment mechanism. In that the aim is to support theoretical development, we deliberately consider a minimally complicated (i.e., maximally stylized) model. Further research is needed to explore the empirical validity of the theoretical assumptions grounding our model. Even so, our findings provide a compelling rationale for the development of specific and explicit theories of evolution within economic systems. Even the most highly stylized of entrepreneurs face a shifting ``opportunity space'' where the number of potential business opportunities is effectively unbounded. This suggests that discrete choice is inherently limited as a framework for studying entrepreneurial growth. The selection mechanism in more realistic future models ought perhaps to be fundamentally evolutionary in nature.

The remainder of this paper is organized as follows. Section~\ref{sec:model} describes the growing network model. Section~\ref{sec:results} presents our findings on its early behavior. Section~\ref{sec:theory} details the theoretical argument relating growth under opportunistic attachment to entrepreneurial growth in economic systems. Section~\ref{sec:discussion} discusses the implications of the findings insofar as the proposed economic theory is sound.

\section{Growing network model} \label{sec:model}

This work uses computational simulation to explore a minimal model of network growth under opportunistic attachment. The initial configuration is, in each instance of the simulation, a simple cycle of three nodes. Incoming nodes then enter the network one by one, each placing a single incoming link and a single outgoing link. The node entering the network at step $N$ encounters an existing network of $N-1$ nodes, where there are $(N-1) \cdot (N-2)$ possible entry points available to select. The entering node considers the $(N-1) \cdot (N-2)$ potential new networks, where it has joined at each of the possible entry points. Our metric for the quality of the ``opportunity'' available to the incoming node at an entry point is the score given to the incoming node when the PageRank algorithm is run on the potential new network. This score reflects the stationary probability of a random walk at the new node's position within the potential new network; we include 5\% probability of random jumps~\parencite[][$\alpha = 0.95$]{page_pagerank_1999}.

The selection probability assigned to each available entry point derives from the $(N-1) \cdot (N-2)$ projected PageRank scores, scaled by the largest value. This vector of scores is raised to a power of $\gamma$ before being divided by its sum, in close analogy to generalized models of network growth under preferential attachment~\parencite[][, see below]{krapivsky_connectivity_2000}. This probability weighting is used to stochastically select the position at which the incoming node enters the network. When $\gamma = 0$, positions are randomly selected. When $\gamma$ is high enough, entering nodes are consistently choosing optimally with respect to the opportunity score. This ``intesity'' of the opportunistic attachment mechanism is used in the remainder of this work to explore the effect of the mechanism on system dynamics, node dynamics, and the resulting network structure. Perhaps it should be explicitly stated that once a node has joined the network, is takes no further actions. 

\subsection{Related literature} \label{sec:model:related}

Growing network models define how incoming nodes join a network to explore the network structure that arises as nodes enter, sequentially. The Barabasi-Albert model is perhaps the most well-known growing network model, where incoming nodes select existing nodes to connect with in proportion to their existing degree~\parencite{barabasi_emergence_1999}. A probability of attachment that is linear with respect to degree gives rise to networks with a power-law degree distribution, as early nodes experience cumulative advantage. Growing network models that include preferential attachment have been developed for use with networks of businesses~\parencite{atalay_network_2011,ozaki_integration_2024} and extended to incorporate elements such as homophily and triadic closure~\parencite{bramoulle_homophily_2012}. Isolating such mechanisms can be especially insightful in that relatively small differences at the node level can lead to major differences in network structure, over time~\parencite{asikainen_cumulative_2020}.

In the generalized version of the Barabasi-Albert model, incoming nodes select existing nodes to connect to in proportion to their degree raised to a power of $\gamma$~\parencite{krapivsky_connectivity_2000}. When $\gamma = 0$, each incoming node is choosing randomly and early nodes experience a steady advantage. When $\gamma$ is high, entering nodes consistently select the highest-degree nodes and the network structure degenerates to hub-and-spoke. Linear dependence of the probability of attachment on degree ($\gamma = 1$) is special, as this is a reasonable approximation for systems where more links make a node more visible. This has been found to apply on social media, in bibliometrics, in systems driven by referrals, and in systems where incoming nodes copy the connections of existing nodes~\parencite{kleinberg_web_1999,kumar_stochastic_2000,salganik_experimental_2006,pham_pafit_2015,pham_joint_2016}. 

This model can be further generalized into a \emph{family} of growing network models. Within a discrete choice framework, incoming nodes can be modeled to select among the existing nodes based on a preference for \emph{any} features of those nodes~\parencite{overgoor_choosing_2019}. Degree is just one possible preference metric by which incoming nodes select among existing nodes. In this sense, the Barabasi-Albert model could be considered the minimal model of a network growing under ``preferential attachment''.

The model presented in this work would be part of a family of growing network models where incoming nodes select among the network positions available to them, based on features of that \emph{available entry point}. This reflects scenarios where the incoming node is evaluating, not the nodes it would connect to, but, the role it would step into. Let us now envision that there may be scenarios where a single metric, analogous to degree, might be useful in approximating the benefit to an incoming node of entering the network at a particular entry point; we select projected PageRank as the opportunity metric for our minimal model of a network growing under ``opportunistic attachment''.

\section{Results} \label{sec:results}

\begin{figure}[ht]
  \centering
  \includegraphics[width=1.0\textwidth]{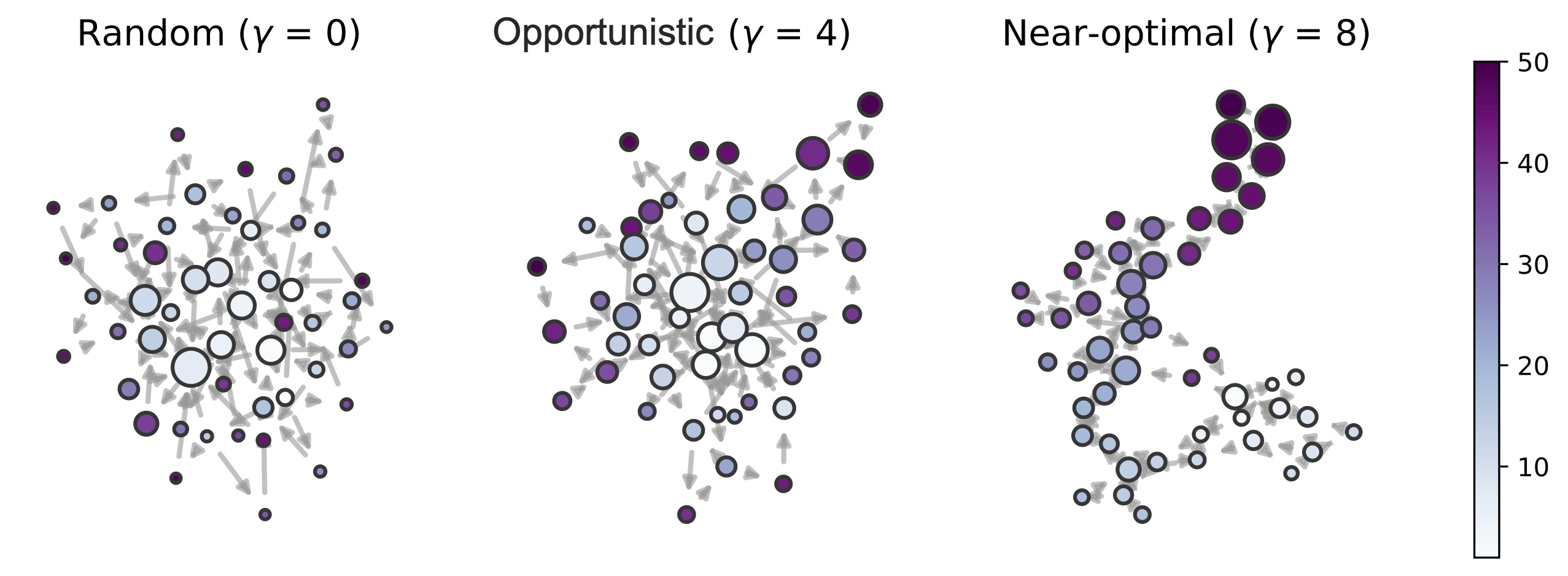}
  \caption{\textbf{Growing networks of 50 nodes.} Visualized structure of growing networks where attachment was random, opportunistic, or near-optimal with respect to projected PageRank ($\alpha=0.95$, $\gamma=[0,4,8]$). Nodes are sized by realized PageRank and colored by entry time; the initial nodes are the lightest.}
  \label{fig:networks}
\end{figure}

Network growth under opportunistic attachment gives rise to complex dynamics characterized by an endogenous opportunity space, path-dependence, and turnover in node rank. Figure~\ref{fig:networks} shows three snapshots of growing networks, at $N=50$, under three different selection intensities. Random selection gives rise to a random network where early nodes see a steady advantage; this is the same network structure as that resulting from a model of network growth under no preferential attachment. Near-optimal selection with respect to the opportunity metric gives rise to networks that grow in a principal direction, occasionally folding back on itself. Opportunistic selection at an intensity between the two extremes produces networks with less predictable, almost organic, patterns of growth.

The remainder of this section describes in greater detail the system dynamics, node dynamics, and network structure that arise from this growing network model. Section~\ref{sec:results:system} considers the distribution of opportunity scores available to incoming nodes and an example of how this introduces path-dependence into the system. Section~\ref{sec:results:nodes} outlines the resulting dynamics of node rank. Finally, Section~\ref{sec:results:runaway} describes the degenerate network structure resulting from optimal-choice opportunistic attachment.

\subsection{Opportunity space} \label{sec:results:system} 

Incoming nodes in our minimal model encounter a precisely defined opportunity space. They can place themselves as ``selling to'' and ``buying from'' any pair of existing nodes, and entry points with higher PageRank are considered better opportunities. The growth process itself changes this opportunity space in two ways. First, the incoming node at step $N$ expands the opportunity space by $N$ additional possibilities. Second, the entry point selected by the incoming node changes the projected PageRank of the existing $(N-1) \cdot (N-2)$ possibilities. In this specific sense, the opportunity space is endogenous.

\begin{figure}[ht]
    \centering
    \includegraphics[width=0.49\textwidth]{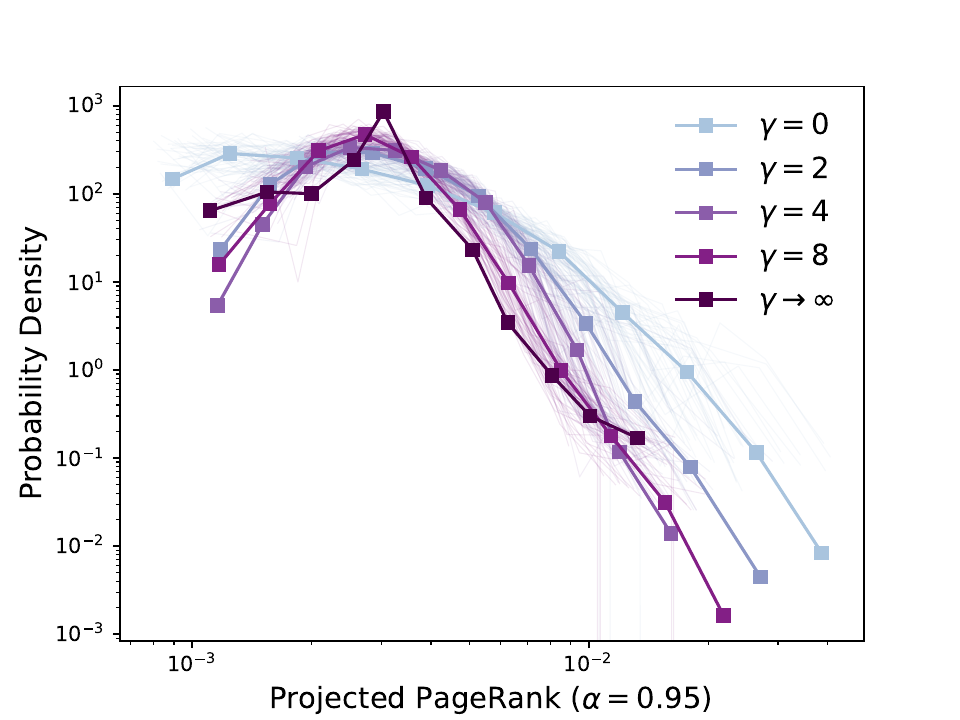}
    \includegraphics[width=0.49\textwidth]{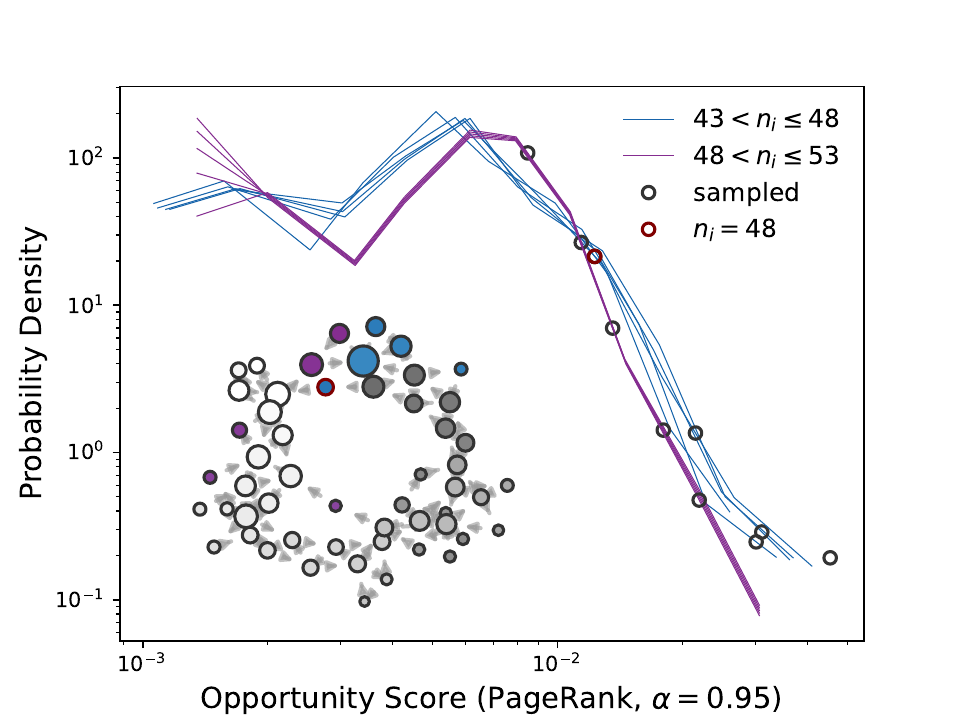}
    \caption{\textbf{Distributions of the opportunity metric.} Left: Distribution of scaled projected PageRank values for the 9702 network positions available to incoming node $100$, within opportunistic attachment models at different intensities. The labeled distributions combine 50 independent realizations, which are also shown separately as fine lines in the same colors. Right: Distributions of the opportunity metric for incoming nodes $44$-$53$ within a specific realization of the growing network model with $\gamma=8$. The selected network positions are noted in a visualization of the growing network upon entry of node $53$, as well as the corresponding values of the expected revenue metric. Node $48$ looped the network; it and its sampled score are circled in red.}
    \label{fig:endogenous}
\end{figure}

As a result, the nodes entering this stylized system under models with different intensities of opportunistic attachment encounter systematically different opportunity spaces. Figure~\ref{fig:endogenous} (left) shows the distribution of the opportunity metric as encountered by node $100$, for different values of the intensity $\gamma$. At lower intensities there are relatively many network positions available, at this point in the growth process, with a score substantially higher than most. Especially good opportunities become more scarce with even mildly opportunistic attachment. Higher intensities compress the range of opportunities further. 

Near-optimal opportunistic attachment on this minimal model gives us particular insight into how this happens. Figure~\ref{fig:endogenous} (right) shows distributions of the opportunity metric for a specific realization of the growing network model, with $\gamma=8$. In this realization, node $48$ selects a network position that appreciably shifts the network structure and thus also the opportunity space. Nodes entering before node $48$ see wider distributions of the expected revenue metric (in blue) and select highly opportune network positions with a score near the tail. These nodes are colored blue in the network visualization; note that they grew the network along its principal direction. Node $48$ selects a relatively poor opportunity, by chance, and enters at a position that loops the network almost all the way back on itself. This shift in the network structure shifted the opportunity space. Subsequent nodes (in purple) encounter distributions of the opportunity metric with a compressed range. This results in the entry of several nodes at network positions far from the most recent nodes. We conclude that stochasticity is of substantial importance in the time-evolution of the opportunity space, and that this time-evolution becomes jagged under highly opportunistic attachment.

\subsection{Dynamics of node rank} \label{sec:results:nodes}

Although incoming nodes are identical, the opportunities available to them are not. Moreover, some nodes are substantially ``luckier'' in that they happen to choose a more advantageous position given the stochasticity of selection. However, the \emph{expected} PageRank used to construct the opportunity space for an incoming node coincides with the \emph{realized} PageRank only at the step it enters the network. As the network continues to grow, it is the stochastic choices of subsequent nodes that determine the dynamics of node rank over time. 

\begin{figure}[ht]
  \centering
  \includegraphics[width=1.0\textwidth]{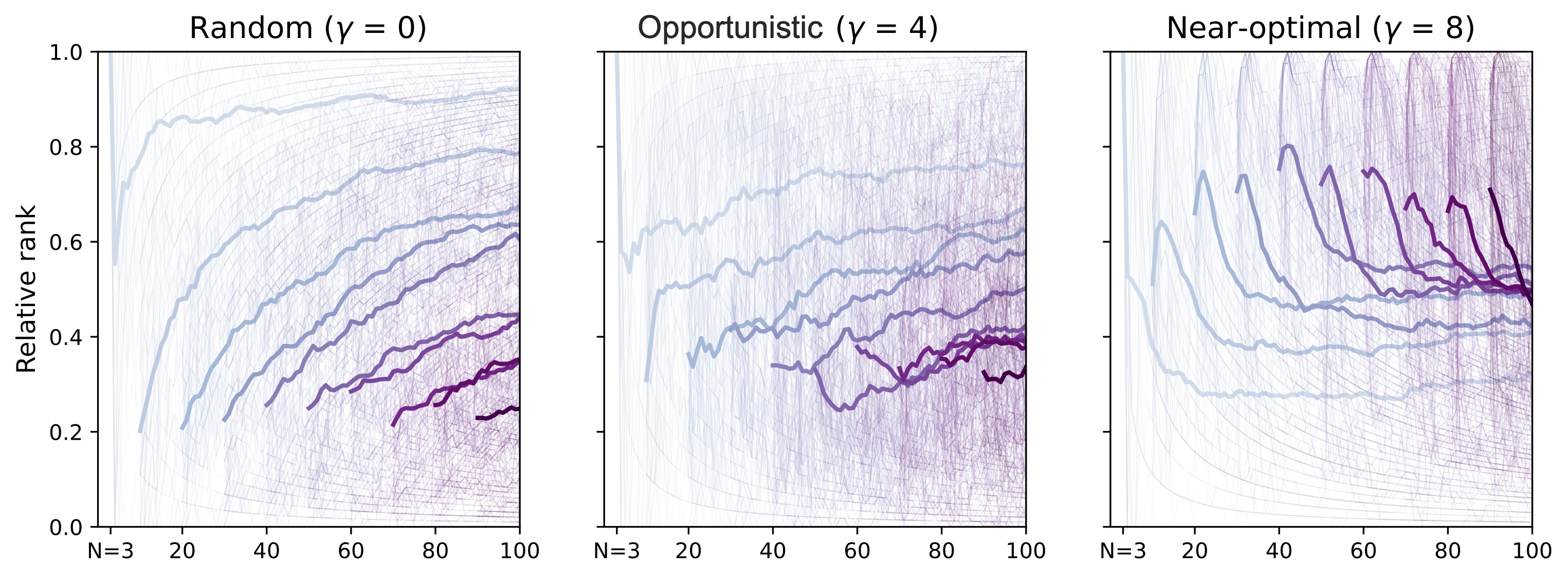}
  \caption{\textbf{Relative rank as networks grow.} Relative rank of nodes within growing networks where attachment was random, opportunistic, or near-optimal with respect to projected PageRank ($\alpha=0.95$, $\gamma=[0,4,8]$). Solid lines track the average rank of every tenth incoming node, plus an initial node, across 50 independent realizations. Fine lines track the relative rank of these same incoming nodes, separately for each realization.}
  \label{fig:node-dynamics}
\end{figure}

Here we consider the dynamics of node rank with respect to \emph{realized} PageRank. Figure~\ref{fig:node-dynamics} shows the relative rank of every tenth incoming node as networks grow to a size of $100$ under three different intensities of the opportunistic attachment mechanism. In a growing network without opportunistic attachment (or, preferential attachment) early nodes are at an advantage and will tend to remain at a higher rank. Random selection thus strongly favors early nodes. The three initial nodes, in particular, will tend to stay above the 80th precentile in rank with respect to realized PageRank. Subsequent nodes will tend to enter at a relatively low rank, while the existing nodes, on average, progressively rise in rank.

As opportunistic attachment increases in intensity, earlier nodes no longer win out as the network continues to grow. Under near-optimal attachment with respect to projected PageRank, later nodes tend to enter at a relatively high rank and newly arrived nodes routinely become the highest-ranked node within a few subsequent steps. However, they are soon overtaken by even newer, highly opportunistic, entrants. Later nodes retain their advantage over those that entered earliest, on average. 

The ability to select an opportune network position introduces a potential benefit to arriving later, but this does not \emph{negate} the benefit of arriving early. Opportunistic attachment at intensities where neither effect dominates, gives rise to, above all, considerable instability in the rank dynamics. The average rank for entrants at a particular $N$, across realizations of the model, is noisier and less representative. ``Luck'' matters most and the modal experience of an incoming node is for their rank to change substantially with the stochastic choices of subsequent incoming nodes.

\subsection{Degenerate case} \label{sec:results:runaway}

\begin{figure}[ht]
  \centering
  \includegraphics[width=\textwidth]{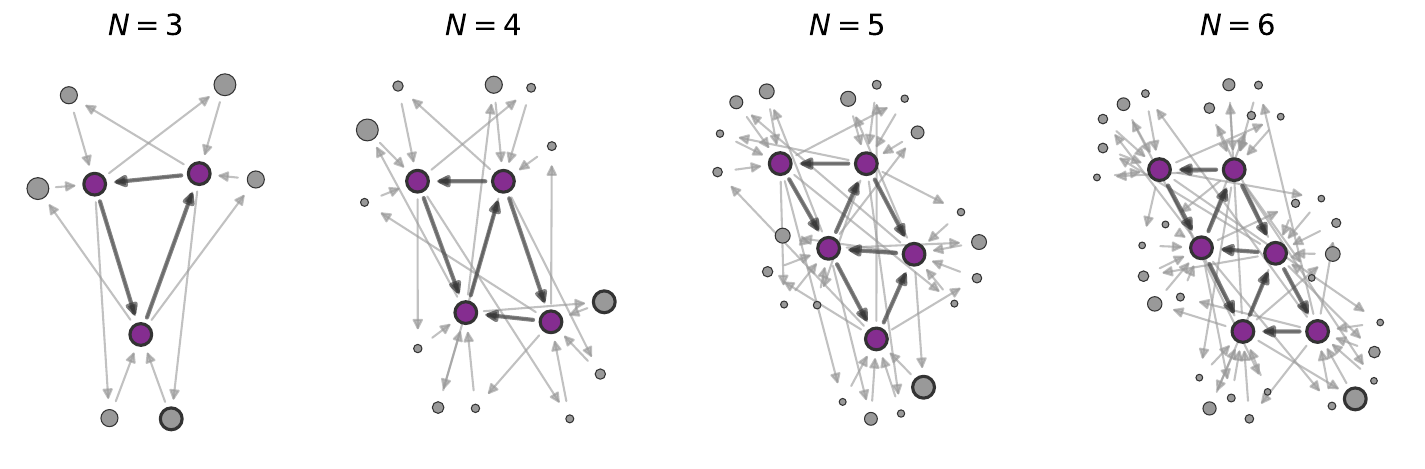}
  \caption{\textbf{Growing network under optimal selection.} Visualized structure of a growing network where attachment was optimal with respect to projected PageRank ($\alpha=0.95$, $\gamma \rightarrow \infty$). Existing nodes are colored and selected nodes are circled. Possible entry points are sized by projected PageRank with $\gamma = 4$.}
  \label{fig:optimal}
\end{figure}

Stochasticity is needed for growing networks to maintain a rich structure. Under preferential attachment with respect to degree, incoming nodes selecting \emph{optimally} would all connect to the same initial node(s). This degenerate case results deterministically in an extreme hub-and-spoke structure~\parencite{barabasi_emergence_1999}. Under opportunistic attachment with respect to projected PageRank, for this minimal model where each node incoming places one in-link and one out-link, optimal selection results in a laddered network structure where each incoming node connects deterministically to the previous two (Figure~\ref{fig:optimal}). 

This degenerate case reveals an intriguing additional dynamic. Incoming nodes find the best opportunities around the next-newest nodes. And so, the network will grow inexorably in the principal direction established by the first and only stochastic choice ($N=4$). Notably, this pattern of growth also results in the deterioration of the opportunities around the older nodes. As the best opportunities get progressively farther away from the oldest nodes, the nearby opportunities also get less and less appealing to incoming nodes; one could say that the oldest nodes are ``left behind'' as the network grows. Should this phenomenon be general to growth under opportunistic attachment, it would be highly relevant to systems where growth is endogenous and opportunity begets opportunity.

\begin{figure}[ht]
  \centering
  \includegraphics[width=0.45\textwidth]{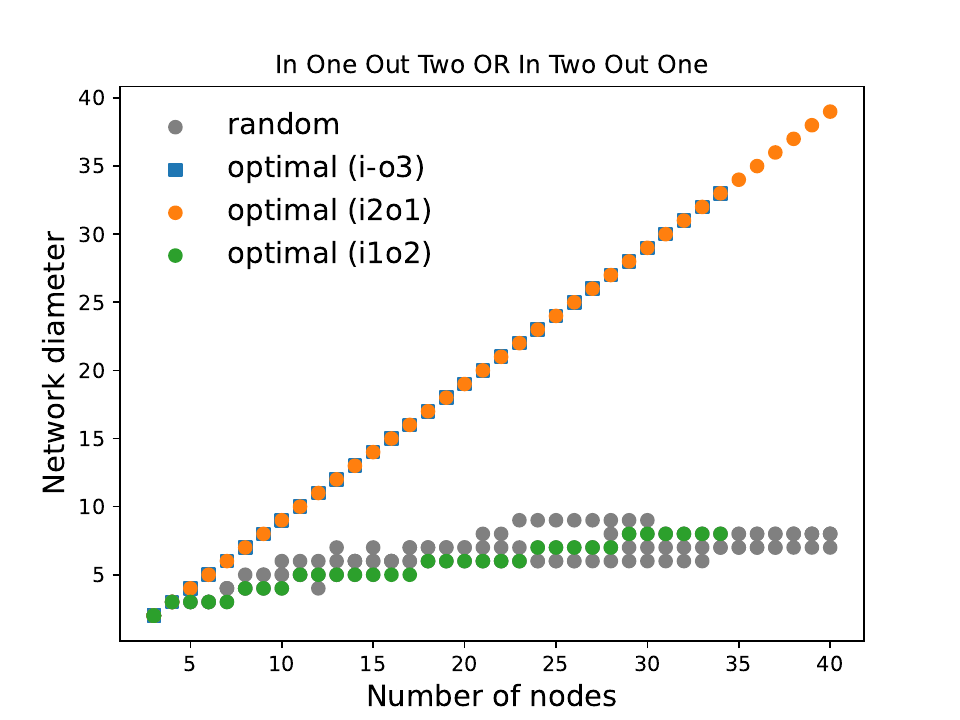}
  \caption{\textbf{Network diameter under optimal selection (i-o3).} Incoming nodes are allowed one in-link, one out-link, and one additional link. The diamter increases linearly, the hallmark of the laddered degenerate network structure, when this link is allowed to be in either derection or constrained to be an extra in-link. In the former case, nodes consistently select an extra in-link. The degenerate structure does not appear when the additional link is constrained to be an extra out-link. Comparison to random selection shown in grey.}
  \label{fig:diameter_i-o3}
\end{figure}

\begin{figure}[ht]
  \centering
  \includegraphics[width=0.45\textwidth]{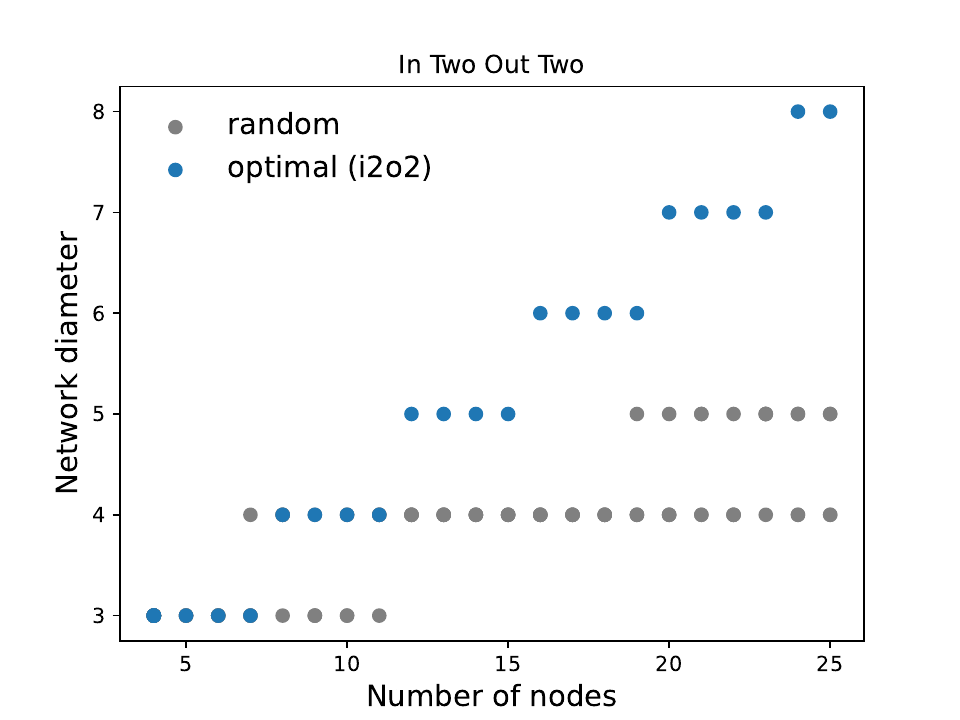}
  \caption{\textbf{Network diameter under optimal selection (i2o2).} Incoming nodes are allowed one in-link, one out-link, and one additional link in each direction. The diameter increases linearly under optimal selection. Comparison to random selection shown in grey.}
  \label{fig:diameter_i2o2}
\end{figure}

Within our minimal model, laddered motifs appear to be emblematic of opportunistic attachment in the same sense that hubs are emblematic of preferential attachment. Recall from Figure~\ref{fig:networks} that near-optimal selection results in networks that grow mostly along a principal direction before folding back on themselves. And, the incoming nodes whose selected position folds the network back on itself are those that make an especially ``unlucky'' stochastic choice (Figure~\ref{fig:endogenous}, right). Shorter ladder motifs appear also in networks growing under less intense opportunistic attachment (Figure~\ref{fig:networks}, middle). This degeneracy persists when incoming nodes are allowed an extra link in either direction or an extra link in each direction; the diameter of the resulting networks increase linearly with the number of nodes (SI Figures~\ref{fig:diameter_i-o3}~\&~\ref{fig:diameter_i2o2}). However, imposing additional constraints can prevent the system from producing this degenerate structure under optimal attachment.

This phenomenon also serves to highlight the limitations of this model. Full exploration of the degenerate case is possible only with the development of further models in the opportunistic attachment family, relaxing some constraints and imposing others. For example, directly specifying the ego-network for incoming nodes is a very strong modeling constraint. On the other hand, exhaustive consideration of all possible entry points is a rather generous perspective to take on the acuity of incoming nodes. Moving forward, one might imagine developing models were selection is \textit{evolutionary} instead of \textit{discrete choice}, in that only the ``luckiest'' nodes survive. For now we conclude only that this unexpected phenomenon---network growth shooting off in one direction---deserves further study in systems where opportunistic attachment may be operating.

\section{Entrepreneurial economies} \label{sec:theory}

Entrepreneurial economies are one domain where opportunistic attachment mechanisms may be operating. To the extent that this is the case, a growing network model fully isolating this mechanism could be a useful tool in the development of relevant theory~\parencite{healy_fuck_2017}. Here we outline a series of dramatic simplifying assumptions that relate the entry of new business in an entrepreneurial economy to our growing network model, providing a theoretical argument for each assumption in turn. First, we represent an economic system as a network of buyer-supplier ties among economic entities (Section~\ref{sec:theory:network}). We then extend a mathematical argument from~\cite{leontief_money-flow_1993} to consider the stationary density of a walk process on such a network as a notion of economic equilibrium (Section~\ref{sec:theory:equilibria}). In precisely this sense, our growing network model represents a sequence of economic equilibria and our opportunity metric can be interpreted as the expected turnover or revenue of a hypothetical new business. 

The result is a \textit{highly stylized} representation of a growing entrepreneurial economy. It is also worth noting that we implicitly employ the Universal Payment System Approximation in representing an entire economic system as a network in this way (Section~\ref{sec:theory:universal}). This helps us articulate some key limitations and points us towards the possibility of exploring these assumptions empirically.

\subsection{Networks of buyer-supplier ties} \label{sec:theory:network}

Input-output linkages can be considered as a network at many levels of granularity. So-called ``production networks'' can be constructed for among a handful of sectors, for among dozens of industries and, ultimately, for among millions of individual firms~\parencite{carvalho_production_2019}. As findings get translated from one extreme to the other, network analysis of firm-level production networks is becoming recognized as the fully disaggregated version of sector-level input-output analysis~\parencite{bacilieri_firm-level_2023}. Most notably: aggregate fluctuations in economic output can arise from the propagation of idiosyncratic productivity shocks over heterogeneous firm-level production networks~\parencite{acemoglu_network_2012,magerman_heterogeneous_2016,carvalho_production_2019}. Modern simulations of, e.g., supply disruption, are increasingly based on networks constructed from empirical data on buyer-supplier ties~\parencite{inoue_firm-level_2019,carvalho_supply_2021}.

The links of our network are economic ties from a buyer to a supplier. In the context of our model, we assume that these ties are durable and that payment for the goods and services of this stylized economy would flow primarily over these ties. For this there is some empirical support, plus theoretical rationales offered by a surprisingly diverse set of intellectual traditions. From a survey of 1501 manufacturing firms in Hungary, Slovakia, and Romania, we know that ties accounting for more than 10\% of sales usually persist for years and often involve product modification or other investments on the part of the supplier~\parencite{bekes_supplier-buyer_2021}. \cite{uzzi_social_1997} combines ethnographic fieldwork with systematic network data to map the garment industry of New York City; while one-off transactions across market relationships are found to be prevalent, a disproportionately large share of business goes to established ties. That an economic system is made up of such ties, to a first approximation, is consistent with network theory of organization~\parencite{powell_neither_1989,white_markets_2004}, relational contracting~\parencite{baker_relational_2002}, and diffusion~\parencite{perra_random_2012}.

Note that nodes of our network are economic entities, and we avoid making distinctions between households, firms, and other specific categories of entity. This is a deliberate simplification with respect to common practice in agent-based modelling of economic systems~\parencite{delli_gatti_macroeconomics_2011,mankiw_macroeconomics_2015,riccetti_agent_2015}. Distinctions between households, firms, banks, and other categories of economic actors are required to model markets as matching processes, and they are certainly valid within the context of a legal infrastructure that enforces them. However, they may or may not be inherent to specific economic phenomena. Indeed, the literature on networks within development economics often highlights the \emph{lack} of a clear distinction between firms and households in many situations~\parencite{fafchamps_trade_1997,lyon_trust_2000,jackson_social_2012}. 

\subsection{Economic equilibrium as the stationary density of a walk process} \label{sec:theory:equilibria}

The notion of equilibrium in an economic system is related to the steady-state of a dynamical process. In \cite{leontief_money-flow_1993}, the authors show that ``the flow of money [can be] described as a Markov chain. Its ergodic state is equivalent to the economic equilibrium.'' Their Markov chain, which we will call $T$, is derived from a 1985 Hungarian input-output table with 5 sectors in the following way: first, entries are normalized to represent the inputs to a sector's production as a proportion. Then, the transpose of this table is read as a matrix of transition probabilities for money among the sectors. In its simplest form, where unit velocity of money is assumed everywhere, the process described by $T$ will converge to its steady-state equilibrium so long as the system allows an equilibrium (\textit{Theorem 1}, p. 228). This result is described as follows: 
\begin{quote}
    In economic terms, undisturbed circulation of money may secure equilibrium. NB: no further assumption, besides an abstract numerical possibility of equilibrium, has been used here and this abstract possibility can be expressed by a very simple equation: $det(1 - T) = 0$ \parencite{leonfief_structure_1941}. No terms loaned from labour theory nor any notions of marginalism entered our exposition. Equilibrium prices and proportions, together with their practical attainability, can be based on and demonstrated by simple money-flow considerations.~\parencite[][p. 228]{leontief_money-flow_1993}
\end{quote}

At higher granularity, this translates to the stationary density of a walk process over a network of buyer-supplier ties among economic entities. Unlike the production functions typical of standard input-output analysis, the mechanics of money at the level of individual entities need not be modeled differently for different sectors. Here, the ``Households'' and ``Governments'' sectors of the early input-output table in \cite{leontief_money-flow_1993} are treated no differently from the ``Extraction'', ``Manufacture'', or ``Commerce'' sectors. Specifically, let $m_i$ denote the amount of money in the till of sector $i$, let $\mathbf{m}$ denote the vector of these values, and let $\mathbf{p} = \mathbf{m} \div \| \mathbf{m} \|$ denote the normalized vector. Equation~\ref{eqn:DTRW} then reproduces the matrix expression for the steady-state $\mathbf p^{*}$ of the Markov chain $T$ in the simplest model where money is assumed to move at unit velocity (\textit{Theorem 1}, p. 228). The very same expression gives the stationary density of a discrete-time random walk over a finite network of $N$ nodes where $T$ is the Markov chain representing the transition probabilities among the nodes~\parencite[][Eq. 30]{masuda_random_2017}. Specifically, the transition probability matrix $T$ is a normalized version of the $N \times N$ adjacency matrix $A$ where each entry $A_{ij}$ is divided by the (possibly weighted) out-degree of node $i$~\parencite[][Eqs. 24-25]{masuda_random_2017}. 
\begin{equation} \label{eqn:DTRW}
    \mathbf p^{*} = \mathbf p^{*} T 
\end{equation}

The stationary density $\mathbf p^{*}$ is a unique ergodic equilibrium state when the directed network $A$ is strongly connected~\parencite{newman_networks_2010,masuda_random_2017}. In practice, this condition is often ensured by modifying the transition probability matrix $T$. PageRank introduces random jumps into the random walk by combining $T$ with a preference vector $u$. In the standard version of the algorithm, $T$ is modified such that node $i$ sees its network transition probabilities $T_{i}$ with probability $\alpha$ and uniform transition probabilities $u_{i} = 1/N$ with probability $(1 - \alpha)$; nodes with no out-degree see only $u_{i}$~\parencite[][Sec. 5.2.1]{masuda_random_2017}. In this work we use $\alpha = 0.95$, which would correspond to a stylized scenario where 5\% of expenditures occur across ties with random entities.

This specific notion of economic equilibrium engages three inter-related simplifications: an unweighted network, Markovian transitions, and discrete time. Networks of buyer-supplier ties are often represented as weighted networks in that total transaction volumes across ties are highly heterogeneous~\parencite{mattsson_circulation_2023,bacilieri_firm-level_2023}. However, total transaction volumes across ties are ultimately an observable \emph{outcome} of the dynamical process. Depending on the model of the process, attaining this outcome may or may not require the input network underlying the circulation of money to be represented as a weighted network. Non-linear money transport models are used to approximate total transaction volumes on unweighted firm-firm networks~\parencite{tamura_diffusion-localization_2018,ozaki_integration_2024}. Moreover, payments are also weighted and fully granular models of transaction processes would be represented in continuous time~\parencite{mattsson_trajectories_2021}. Equation~\eqref{eqn:DTRW} is not a \emph{realistic} representation of the circulation of money over a network of buyer-supplier ties. A discrete-time random walk is simply the \emph{minimum viable model} for undisturbed circulation of money that would give rise to an economic equilibrium as per \cite{leontief_money-flow_1993}.

\subsection{Universal payment system approximation} \label{sec:theory:universal}

Our stylized economic system implicitly operates within a single, universal, payment system. This is a simplification, in that actual monetary infrastructures are complicated systems of systems~\parencite{james_drafts_2010,galbiati_clearing_2012,mbiti_mobile_2015}. However, direct violations of the universal payment system approximation---such as the 2023 run on deposits entrusted to Silicon Valley Bank---are as spectacular as they are rare~\parencite{cipriani_tracing_2024}. Moreover, approximating a ``general purpose payment system'' is the actual policy objective of regulators in many countries~\parencite{summers_governance_2014}. The universal payment system approximation is widely used also in economy-wide agent-based macroeconomic models~\parencite[e.g.,][]{caiani_agent_2016}. 

Stating this approximation explicitly lets us articulate what it leaves out. Practically speaking, payment systems can falter during periods of serious disruption~\parencite[e.g., the events of September 11, 2001, see:][]{soramaki_topology_2007,bech_illiquidity_2012} or illiquidity~\parencite{galbiati_agent-based_2011}. Substantively, improvements to payment system integration can improve economic outcomes by lowering transaction costs~\parencite{jack_risk_2014}. However, these factors are deemed to be relatively minor. More consequential is that the universal payment system approximation abstracts away from the actual mechanics of money creation and dissolution, both in theory~\parencite{mehrling_vision_1999} and in practice~\parencite{werner_can_2014,mcleay_money_2014}.

\section{Discussion} \label{sec:discussion}

In this paper, we have found that an opportunistic attachment mechanism can produce highly endogenous, path-dependent patterns of network growth. Moreover, we have argued that such growing network models could prove useful as a tool for the development of theory regarding growth in highly entrepreneurial economies. The main result is to suggest that several complex economic phenomena might be usefully related via the concept of a shifting ``opportunity space'' faced by existing businesses and, especially, by entrepreneurs. In particular, our minimal model can be seen as a proof-of-concept towards more formal models where an economic network and its economic equilibrium meaningfully co-evolve. 

Insofar as the proposed economic theory is sound, opportunistic attachment is a relevant mechanism to consider in modelling the creation of new businesses by entrepreneurs keen to start thriving businesses. However, this is \textit{not} to imply that our growing network model is realistic. In fact, new firm entry is necessarily related also to dynamics of innovation~\parencite{ghiglino_random_2012,carvalho_input_2014} and credit provision~\parencite{werner_can_2014,mcleay_money_2014}. Opportunities for entrepreneurs are limited by many practical constraints not captured as an ego-network~\parencite{cassar_entrepreneur_2006}. Moreover, existing businesses actively search for better suppliers and new customers~\parencite{chaney_network_2014,carvalho_input_2014,krichene_tie-formation_2019,krichene_emergence_2019}. While it is possible to develop agent-based models that can serve as calibrated simulations of real systems~\parencite{monti_learning_2023}, these limitations mean that our model is unlikely to be useful in this regard.

Rather, our aim has been to isolate the opportunistic attachment mechanism via deliberate simplifications and to study its potential implications. Highly stylized models of this sort can be especially insightful in the context of dynamic networks, where relatively small differences in node-level behavior can lead to major changes in network structure; so much so that it has been argued ``individual level mechanisms should not be analyzed separately without considering the dynamics of society as a whole''~\parencite{asikainen_cumulative_2020}. Strong path-dependence, for example, is a well-documented feature of real economic systems as a whole\parencite{arthur_complexity_2014,nunn_historical_2007,nunn_long-term_2007,redding_history_2011,thurner_introduction_2018}. This makes opportunistic attachment an interesting candidate mechanism for relating the network structure of an economic system to the direction of its future growth, alongside other known mechanisms including preferential attachment, technological improvement, and splits/mergers~\parencite{atalay_network_2011,mcnerney_how_2022,ozaki_integration_2024}.

Further research is needed to generalize any specific findings beyond small networks, and as our simplifying assumptions are relaxed. While it is uncommon to formally consider the space of possible entry points for incoming nodes, it is possible to bound the effect of arbitrary node addition on some spectral properties of graphs~\parencite{torres_perron_2021}. More readily amenable to mathematical formalism may be business \emph{adaptation} rather than business \emph{entry} in highly entrepreneurial economies. \cite{holme_dynamics_2006} describe a remarkable model where a fixed population of agents aim to achieve higher closeness centrality while maintaining low degree; this setup includes ``the computation of centralities which can be seen as a (fast) dynamical process on the network... similar to other adaptive network models with timescale separation''~\parencite[][p. 34]{berner_adaptive_2023}. It is our fervent hope that the theoretical contribution of this work, relating the stationary density of a random walk on a network of economic entities the economic equilibrium (Section~\ref{sec:theory:equilibria}, Eq.~\eqref{eqn:DTRW}), might motivate the development of adaptive dynamical network models relevant to economic systems.

In the other direction, our results provide a compelling rationale for the development of specific and explicit theories of evolution within economic systems. Our minimal model highlights the extreme limitations of discrete choice as a framework for studying growth in economic systems; actual entrepreneurs face a shifting ``opportunity space'' where the number of potential business opportunities is effectively unbounded. From an evolutionary perspective, this boundlessness becomes a feature rather than a bug. As in evolutionary models of stylized stock markets~\parencite{farmer_market_2002} or (social) business connections~\parencite{holme_dynamics_2006}, there may be simple heuristic strategies for businesses or entrepreneurs that are especially robust to the strategies of others. The most important \emph{selection} to happen within economic systems may be about survival rather than about utility~\parencite{ozaki_integration_2024}. Indeed, Minsky’s ``survival constraint'' has been used to convey the logic that all economic entities must bring in as much as they spend in order to continue participating in an economic system~\parencite{mehrling_vision_1999}.

\subsection*{Acknowledgments} Thank you to Brennan Klein for fruitful discussions and to Marco Pangallo for insightful feedback. Thank you to Thomas Mattsson for initial inspiration, to James McNerney for valuable references, to Christoph Riedl for specific comments, and to Frank Takes for his skill in naming things. Thanks also to Francois Lafond for constructive editorial feedback. The author acknowledges support by the National Science Foundation Graduate Research Fellowship, Grant No. 1451070.

\subsection*{Software \& Data availability}
No data is used in this work. Replication code will be made publicly available via GitHub.

\subsection*{Competing Interests}
The author declares no competing interests.

\printbibliography

\section*{Supplementary Information} \label{sec:appendix}

\begin{figure}[ht]
  \centering
  \includegraphics[width=0.45\textwidth]{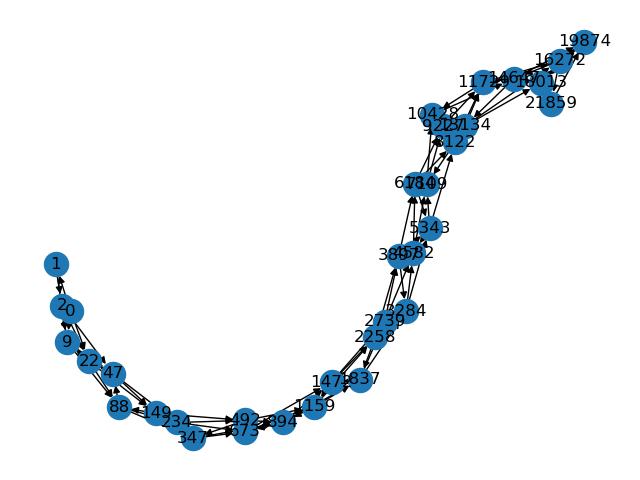}
  \caption{\textbf{Growing network under optimal selection (i-o3).} Incoming nodes are allowed one in-link, one out-link, and one additional link in either direction. This results in a laddered degenerate network structure under optimal selection. The numbers on the nodes gives the number of potential entry points considered in the selection of this position; this grows rapidly.}
  \label{fig:optimal_i-o3}
\end{figure} 

\begin{figure}[ht]
  \centering
  \includegraphics[width=0.45\textwidth]{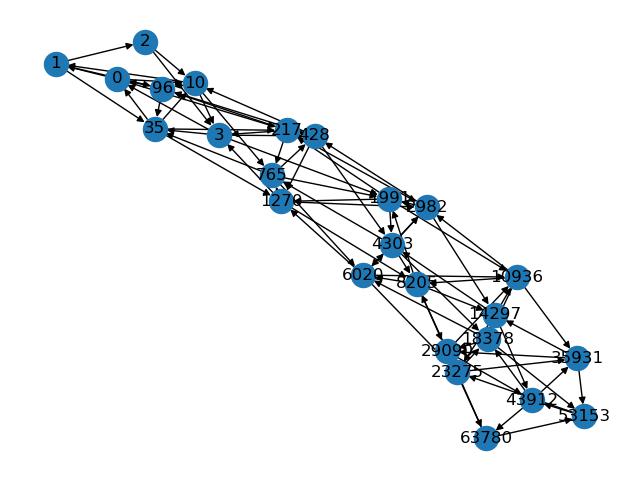}
  \caption{\textbf{Growing network under optimal selection (i2o2).} Incoming nodes are allowed one in-link, one out-link, and one additional link in each direction. This results in a laddered degenerate network structure under optimal selection. The numbers on the nodes gives the number of potential entry points considered in the selection of this position; this grows rapidly.}
  \label{fig:optimal_i2o2}
\end{figure}

\end{document}